\documentclass[aps,prl,twocolumn,showpacs,superscriptaddress,groupedaddress]{revtex4-1}  
\usepackage{graphicx}  
\usepackage{dcolumn}   
\usepackage{bm}        
\usepackage{amssymb}   
\usepackage{amsmath}
\usepackage{xspace}
\usepackage[caption = false]{subfig}
\usepackage{ulem}

\usepackage{times}
\usepackage{color}
\usepackage{ulem}

\newcommand{\swift}{\textit{Swift}}

\newcommand{\gsi}{\,\raisebox{-0.13cm}{$\stackrel{\textstyle>}
{\textstyle\sim}$}\,}

\newcommand{\be}{\begin{equation}}
\newcommand{\ee}{\end{equation}}

\begin{document} 
\title{Tidal disruption jets as the source of \\Ultra-High Energy Cosmic Rays}


\author
{Glennys R. Farrar $^{1}$ and Tsvi Piran $^{2}$}
\affiliation{$^{1}$ Center for Cosmology and Particle Physics,
Department of Physics,
New York University, NY, NY 10003, USA \\  
$^{2}$ Racah Institute of Physics, The Hebrew University of Jerusalem, Jerusalem 91904, Israel }
\date{\today}

\begin{abstract}
{\bf Abstract:}  
Observations of the spectacular, blazar-like tidal disruption event (TDE) candidates  \swift\ J1644+57 and J2058+05 show that the conditions required for accelerating protons to $10^{20}$ eV appear to be realized in the outer jet, and possibly in the inner jet as well.  Direct and indirect estimates of the rate of jetted-TDEs, and of the energy they inject, are compatible with the observed flux of ultra-high energy cosmic rays (UHECRs) and the abundance of presently contributing sources.  Thus TDE-jets can be a major source of UHECRs, even compabile with a pure proton composition.   

\end{abstract}
\maketitle

\section{Introduction}
\label{sec:intro}

More than fifty years have passed since the first detection of a cosmic ray with an energy $\gsi 10^{20}${\rm eV} \cite{Linsley}, yet the nature of UHECRs and the identity of their sources remain a mystery.  
The nature of UHECRs sources depends on their composition, and we focus here on sources capable of producing UHE protons.
Difficulties with the leading source candidates, AGNs and GRBs,  led Farrar and Gruzinov  \cite{fg09}  
to propose that  UHECRs, if protons, must be produced in a new class of powerful AGN transients, as could arise from the tidal disruption of a star or an extreme accretion disk instability around a supermassive black hole at  a galactic center.  
Following the detection of a relativistic outflow from the  tidal disruption event (TDE) \swift\ J1644+57 \cite{Burrows+11,Bloom+11,Zauderer+11}
and from  
\swift\ J2058+05 \cite{Cenko+12} we confront here the viability of a scenario in which UHECRs are protons produced in the jets of tidal disruption events (TDEs).   We begin by recalling the requirements for UHECR acceleration.  Then, we use observations and modeling of \swift\ J1644+57, 
a likely example of a TDE jet seen in ``blazar-mode'' (i.e., looking down the axis of the jet), to test whether individual TDE jets satisfy the Hillas criterion necessary for accelerating protons to  $10^{20}${\rm eV}.   Finally, using the recently measured TDE rate \cite{vfRate14}, we examine whether  TDEs  can account for the observed UHECR energy injection rate and whether they provide a sufficiently large number of active sources to explain the lack of strong clustering in the arrival direction distribution \cite{augerSrcDen13}.

\section{Conditions on Sources of UHECRs}
In order for a CR to be confined during the acceleration process, its Larmor radius must remain smaller than the size, $R$, of the accelerating system.  This places a strict lower bound on $B R$ for UHECR acceleration known as the Hillas criterion, valid for any acceleration mechanism that involves magnetic fields \cite{hillas84}:  
\begin{equation}\label{conf}
B R\gsi 3\times 10^{17} ~ \Gamma ^{-1} \, Z^{-1} \, E_{20} ~ {\rm Gauss \,cm},
\end{equation}
where $B$ is the magnetic field, $\Gamma$ is the bulk Lorentz factor of the jet,
 $Z$ is the charge of the UHECR, and $E$ the CR's energy with $E_{20} \equiv E/ 10^{20} {\rm eV}$.  Eq. (\ref{conf}) implies a lower bound on the total Poynting luminosity required to accelerate protons to UHE, for which the total bolometric luminosity can be taken as a surrogate \cite{fg09}:
\begin{equation}\label{lumi}
 L_{\rm bol} 
 \approx {1\over 6}c\, \Gamma ^4 \, (B R)^2 \, \gsi \, 10^{45}\, \Gamma ^2\, (E_{20}/Z)^2 \, {\rm erg/s}.
\end{equation}
It was shown in \cite{fg09} that if the conditions Eqs. (\ref{conf},\ref{lumi}) are met in an AGN-like jet,  cooling and interaction with photons prior to escape from the accelerator are not the limiting factors in the maximum energy. The strong dependence on $Z$ of  Eqs. (\ref{conf},\ref{lumi}) indicates that the constraints on the UHECR sources are very different for protons or nuclei:  for protons the sources must be amongst the most luminous known EM sources, while for nuclei the requirements are much more modest. 

Denoting the energy injection rate of UHECRs in the range $10^{18}-10^{20}$ eV by ${\dot E}_{_{UCR}} \equiv {\dot E}_{44} \,\, 10^{44} \,{\rm erg} \, {\rm Mpc}^{-3} \, {\rm yr}^{-1}$, for continuous sources the density of sources implied by Eq. (\ref{lumi}) is 
\be  \label{num}
n_{\rm src} \approx 3 \times 10^{-9}  \frac{{\dot E}_{44}}{\epsilon_{_{UCR}} \, \Gamma ^2\, (E_{20}/Z)^2} ~ {\rm Mpc}^{-3},
\ee
where $\epsilon_{_{UCR}}\equiv L_{_{UCR}}/L_{\rm bol}$ is the luminosity in UHECRs relative to the Poynting luminosity.   From the observed UHECR spectrum, \cite{katz+GRB-UHECR09} estimates ${\dot E}_{44} = 2.3$ to 4.5 for source spectra $\sim E^{-2} \, {\rm to} \, E^{-2.5}$, with O(1) uncertainty.  Thus one continuous source with  $\epsilon_{_{UCR}} \sim 1$ within the GZK distance ($\approx 200$ Mpc) would be sufficient to produce the entire observed flux. 
However,  the lack of clustering \cite{augerSrcDen13} in the arrival directions of UHECRs with energies above 70 EeV, implies a constraint on the density of sources whose stringency depends on the characteristic maximum deflections of the UHECRs: for $ 30^{\circ} $ ($3^{\circ}$), the source density must be greater than 
\be \label{nlimsrc}
n_{\rm min} = 2 \times 10^{-5} ~ ( 7 \times 10^{-4}) ~ {\rm Mpc}^{-3}. 
\ee
Thus efficient continuous protonic sources are incompatible with the source abundance requirement 
\cite{fg09,WL09,MuraseTakami09}. Note that  if  deflections in the Galactic and intergalactic magnetic fields are so strong that UHECR arrival directions do not reflect the direction of their sources, the bound Eq. (\ref{nlimsrc}) does not apply. However in this case, the observed correlation \cite{TAaniso14,augerAniso14} with local structure would not be explained.

Eq. (\ref{num}) can be reconciled with the observed minimum source density, Eq. (\ref{nlimsrc}),  if UHECR production is very inefficient with $\epsilon_{_{UCR}} \ll 1$.   However,  inefficiency is not a solution, since even the weakest bound in Eq. (\ref{nlimsrc}) requires $\approx 700$ sources within the GZK distance, whereas powerful steady sources (of any kind) with luminosity larger than $10^{45}$ erg/s are rare \cite{fg09}.  Another way out is via an  acceleration mechanism   which does not involve  magnetic field confinement and thus evades the luminosity requirement Eq. (\ref{lumi}). However an efficient mechanism of this kind is not known. 


Transient sources can evade the previous conundrum.  They must satisfy the Hillas confinement condition embodied in Eqs. (\ref{conf}) and (\ref{lumi}), and furthermore the energy injection condition sets a limit on the energy that must be released in UHECRs in a single transient event:
\be
\label{EUCR}
{\cal E}_{_{UCR}} \equiv {\dot E}_{_{UCR}} \, / \, \Gamma_{_{\rm UCRtran}} =  10^{51}\,{\rm erg} ~ {\dot{E}}_{44}  ~ \Gamma_{_{\rm UCRtran, -7}}^{-1},
\ee  
where $\Gamma_{_{\rm UCRtran}} \equiv \Gamma_{_{\rm UCRtran, -7}} \, 10^{-7}$ Mpc$^{-3}$ yr$^{-1}$ is the rate that the UHECR-producing transients take place.  In addition, the number of sources contributing at a given time must be large enough.  Deflections in the extragalactic magnetic field spread out the arrival times of UHECRs from an individual transient event. 
In the approximation that the deflections are small and many, the resultant characteristic arrival time spread is \cite{waxmanME}:
\be
\label{tau}
\tau \approx \frac{ D^{2} Z^2 \langle B^2 \lambda \rangle}{9 E^{2}} = 
3 \times 10^{5}\, {\rm yr} 
\left(\frac{D_{100}  \, B_{\rm nG}}{E_{20}/Z}\right)^{2}   \lambda_{\rm Mpc}, 
\ee
where $D_{100}$ is the distance of the source divided by 100 Mpc, $B_{\rm nG}$ is the rms random extragalactic magnetic field strength in $nG$, and $\lambda_{\rm Mpc}$ is its coherence length in Mpc.   To be compatible with the Auger bound, 
the number density $n_{\rm eff}$ of contributing transient sources at a given time, must satisfy
\be
\label{neff}
n_{\rm eff} \approx \tau_{5} ~ \Gamma_{_{\rm UCRtran, -7}} \, 10^{-2}\,{\rm Mpc}^{-3} \, \geq n_{\rm min} = 2 \times 10^{-5} ~ {\rm Mpc}^{-3},
\ee
where $ \tau_{5} \,10^{5}\, {\rm yr}$ is the mean time delay averaged over sources.
 
In order for any transient UHECR source type to be viable it must therefore satisfy three requirements, the Hillas condition (Eqs. (\ref{conf}) or (\ref{lumi})) and also Eqs. (\ref{EUCR}) and (\ref{neff}).  
The classical transient source candidate, GRBs,  \cite{waxman95} satisfy easily the first and third conditions, but 
there is debate whether their energy output is sufficient to satisfy the second condition, Eq. (\ref{EUCR}), unless their UHECR energy output exceeds significantly their photon output \cite{fg09,eichler+GRB-UHECR10,katz+GRB-UHECR09}.  This, and the lack of the expected high energy neutrinos \cite{IceCubeNatureGRBs12}, makes  GRBs less favorable source candidates.

A number of  TDE flare candidates have been detected and followed up in real-time: \swift\  J1644+57  \cite{Burrows+11,Bloom+11,Zauderer+11},  PS1-10jh \cite{gezari+Nature12} and PS1-11af \cite{chornock+14}.  Two candidates had been found earlier in archival SDSS data \cite{vf11} and one was subsequently found in archival Swift data  \swift\ J2058.4+0516 \cite{cenkoSw2058}.  (Still earlier TDE candidates were put forward in \cite{donley02,Gezari08}, but AGN-flare background could not be well characterized for those observations, so the origins of the flares were uncertain.) 
 
\section{TDEs as UHECR sources}

In this section we address whether TDE jets are good source candidates for the protonic UHECR scenario.
We begin with the Hillas  condition, Eq. (\ref{conf}), which must be satisfied for any UHECR source that is based on EM acceleration. Here the recent observations of  \swift\ J1644+57 provide us with an example of a TDE jet with very good multi-wavelength follow-up, enabling the Hillas criterion (Eq. \ref{conf}) to be directly checked.

\subsection{\swift\  J1644+57 and the Hillas criterion}

\swift\ J1644+57 was detected on March 25th 2011 by the {\it Swift} satellite.
Its location  at the nucleus of an inactive galaxy made it immediately a strong TDE candidate.  It  is uniquely suitable for testing whether TDE jets can satisfy the Hillas condition, Eq. (\ref{conf}), because of the thorough multi-wavelength monitoring from its inception for more than 600 days, which has enabled detailed modeling of the conditions in its jet.  
The observations  \cite{Burrows+11,Bloom+11,Zauderer+11} revealed  two different
emission sites:  an inner emission region where the X-rays are emitted, and an outer region
where the radio emission is produced. 

 Basic models for the TDE emission follow the ideas of a gamma-ray bursts, c.f.,  e.g. 
\cite{piran04}:  the central engine of the TDE produces a pair of relativistic jets via the
Blandford-Znajeck \cite{bz77} process, with internal dissipation shocks within the jets accelerating
particles and producing the X-ray emission at relatively short distances from the central engine.
At larger distances, the outflow interacts with the surrounding matter;  this slows it down and produces the radio emission, in an afterglow-like manner \cite{Zauderer+11}.   In the following we consider both the X-ray and radio-emitting regions as possible sites for UHECR acceleration. 
We discuss first the radio emitting region where the situation is clearer, as we can use simple equipartition arguments. We then discuss the situation within the X-ray emitting jet. 

\begin{figure}[b!]
\centering
\includegraphics[width=0.45\textwidth]{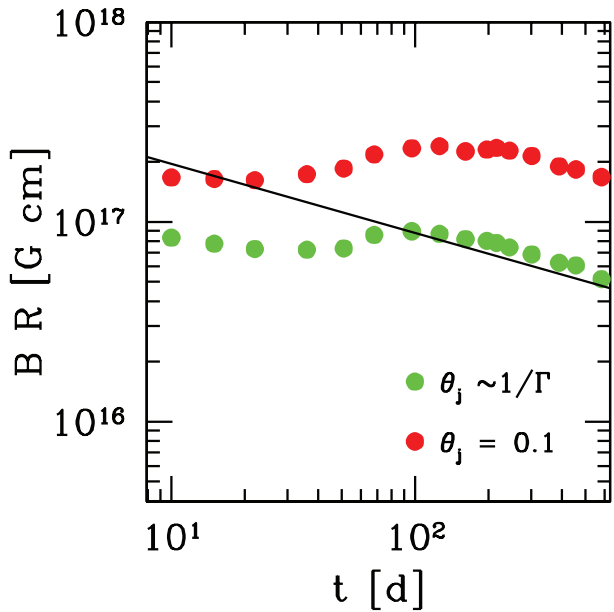}
\caption{ The equipartition values of  $ B R  $  as a function of time, for different relativistic  models for the TDE \swift\ J1644+57:  a narrow jet with an opening angle $0.1$ (red dots), a wide jet  (green dots) and the detailed model of \cite{Kumar+13} (solid line). }  \label{fig:J1644_radio} 
\end{figure}

The conditions within the radio emitting region have been analyzed using relativistic equipartition considerations (see the Appendix). In the relativistic regime the equipartition analysis depends on the geometry of the emitting regions and  there are two possible solutions  \cite{Barniol+13b} shown in  Fig. \ref{fig:J1644_radio}.
For the most reasonable geometry -- a narrow jet with an opening angle 0.1 -- the equipartition value of $B R$   is slightly larger than $10^{17}$ Gauss cm.   Given 
that the equipartition estimates  
yield a lower limit on the energy, this can be considered compatible with the Hillas condition.  A more detailed model that takes into account the inter-relation between the X-ray jet and the radio emitting electrons  \cite{Kumar+13} yields  $B R$  larger by a factor of a few  than the simple equipartition estimate 
\cite{Barniol+13b} (see  Fig. \ref{fig:J1644_radio});  at early times $B R \approx 3 \times 10^{17}$ Gauss cm.  
Thus,  the outer radio emitting region of the \swift\ J1644+57 TDE jet appears likely to have had conditions for UHECR acceleration. 

During the first two days, the X-ray luminosity of \swift\ J1644+57  fluctuated reaching isotropic equivalent peak luminosities of  $\approx 10^{48}$ ergs/sec.  Models of the X-ray emission  as arising from a relativistic blob of plasma \cite{Burrows+11} yield 
estimates of $B R$ ranging from a few $\times 10^{15}$  to $5 \times 10^{17}$ Gauss cm, depending on 
 the dominant energy of the jet (Poynting flux or baryonic), the emission mechanism (synchrotron or external IC), the position of the emission region and the relative contributions of the disk and the jet. 
It is possible but not certain that the X-ray emitting regions in \swift\ J1644+57 also had conditions for UHECR acceleration.  If so, from this point of view TDEs resemble powerful AGNs in the way they satisfy their Hillas condition, with UHECR acceleration being possible in two different regions.   

\subsection{Energy budget and source abundance}
With a total energy of $\approx 10^{54} \,(M_{*}/M_{\odot})$ erg available in a tidal disruption event, only a small fraction -- $ 10^{-3} \, (\langle M_{*}\rangle/M_{\odot})  \, {\dot{E}}_{44} \, \Gamma_{_{\rm UCRtran, -7}}^{-1}$, where $M_{*}$ is the mass of the tidally disrupted star --  needs to go to UHECR production in order to satisfy Eq. (\ref{EUCR}), if the rate of TDEs producing jets capable of accelerating UHECRs is adequate.   

The rate of TDEs in {\it inactive} galaxies has recently been determined based on the discovery of two TDEs in a search of SDSS Stripe 82 \cite{vfRate14}.  In volumetric terms
\be \label{TDErate}
\Gamma_{\rm TDE} =  (0.4-0.8) \cdot 10^{-7\pm 0.4}\,{\rm Mpc}^{-3}\,{\rm yr}^{-1},
\ee
where the statistical uncertainty is in the exponent and the prefactor range reflects the  light curve uncertainty.
This result is roughly consistent with earlier theoretical \cite{WangMerritt04} and observational estimates (within their uncertainties) \cite{Gezari+09}.  However, more refined estimates are needed because not all TDEs have jets and only a fraction of those may be capable of producing UHECRs; in the following we ignore jets weaker than the \swift\ events which satisfy the Hillas criteria.   

An estimate of the UHECR-producing TDE rate is obtained from the observed TDEs with jets,  \swift\ J1644+57 and J2058+05.  Burrows et al. \cite{Burrows+11} estimate that the observation of one event per seven years by \swift\ corresponds to an all sky rate of $0.08-3.9$  events per year up to the detection distance of $z=0.8$ which contains a co-moving volume of $\approx 100\, {\rm Gpc}^{3}$.  Their subsequent archival discovery of J2058+05 increased this rate by a factor of two and reduced somewhat the uncertainty range, leading to an estimated rate of $ \approx 3  \times 10^{-11}$ Mpc$^{-3}$ yr$^{-1}$ of TDE events with jets pointing towards us, or
\be
\label{SwiftJetRate}
\Gamma_{_{\rm UCRtran, -7}} \approx 3 \, f_b^{-1} \, 10^{-4},
\ee
where $f_b$ is the beaming factor.  

In J1644+57, the total EM isotropic equivalent emitted energy in  X-rays was $ 3 \times 10^{53}$ergs \cite{Burrows+11,Bloom+11}; assuming that this is about 1/3 of the total bolometric EM energy, they estimate that the total isotropic equivalent EM energy injection rate is $ \approx 10^{54}$ ergs. Cenko et al. \cite{Cenko+12} estimate a similar total  EM isotropic equivalent energy for J2058+05. 
Since the isotropic equivalent energy is a factor $f_{b}^{-1}$ larger than the true emitted energy, the beaming factor cancels between rate and energy factors, yielding an estimated 
EM energy injection rate of $\approx 3 \times  10^{43}$ erg \, Mpc$^{-3}$ yr$^{-1}$ without relying on knowledge of the beaming factor.  This falls  short of the required energy injection rate for UHECRs, which is $2-4 \times 10^{44}$ erg Mpc$^{-3}$ yr$^{-1}$. Interestingly a similar situation arises when comparing the observed EM GRB flux with the needed UHECR energy injection rate 
\cite{fg09,eichler+GRB-UHECR10} (but see however \cite{katz+GRB-UHECR09}).   The crux question here, is the relation between the emission observed in a particular EM band and the energy production in UHECRs.  The emission mechanisms and relevant particle energies are so different, that it is far from evident how to relate them.

We can also estimate the energy of the jet from the energy needed to produce the radio signal, using equipartition arguments and thus deriving a lower limit on the actual energy.   Different assumptions about conditions within the radio-emitting region lead to different energy estimates. 
From  \cite{Barniol+13b}, we have an estimate of the minimal isotropic equivalent total energy of leptons and magnetic field that is capable of producing the radio emission produced in J1644+57, $\approx  10^{51}$ erg;  including the energy of the accompanying protons increases the estimated minimal energy by a factor $\sim 5$.   Unlike the relativistic inner jet producing the X-rays, the outflow has slowed to being only mildly relativistic with a  low Lorentz factor by the time it produces the observed radio, so that the radio emission is isotropic even when it arises from a jet.  Therefore, the factor $f_b^{-1}$ in the true rate of jetted-TDEs, Eq. (\ref{SwiftJetRate}), does not cancel out and the implied energy injection rate (including the protonic contribution) is of order $ 2 \times 10^{44}  (f_b/10^{-3})^{-1} {\rm erg} \, {\rm Mpc}^{-3}\,{\rm yr}^{-1} $, roughly what is needed as a UHECR flux. 

The above discussion suggests that TDE jets alone could satisfy the needed UHECR injection rate, but underlines the importance of investigating what relationship should be expected between the EM and UHECR spectra and total luminosity. 

The above discussion produced two independent, compatible estimates of the rate of TDEs with jets.  Using the \swift\ observed rate and estimating the beaming factor to be  $f_{b} \approx 10^{-2} - 10^{-3}$ based on the Lorentz factor of $\sim 10-20$ estimated for \swift\ J1644+57 \cite{Burrows+11}, Eq. (\ref{SwiftJetRate}) gives $\Gamma_{_{\rm UCRtran, -7}}^{-1} \gtrsim 3 \cdot 10^{-2}$.  Or, taking the total rate from \cite{vfRate14} and a jet fraction $\sim 0.1-0.2$ based on the fraction of TDEs detected in the radio \cite{vV+Radio13}, gives $ \Gamma_{_{\rm UCRtran, -7}} \lesssim 0.4$.    Using the lower estimate, 
Eq. (\ref{neff}) gives $n_{\rm eff} \approx   3 \, \tau_{5} \,10^{-4}~ {\rm Mpc}^{-3}$, comfortably compatible with the Auger source density limit in the low-deflection scenario, $n_{\rm min} = 2 \times 10^{-5} ~ {\rm Mpc}^{-3}$.

\section*{Composition}

The reader can ask why it is interesting to consider the possibility that UHECRs are predominantly or exclusively protonic, in view of the observed depth-of-shower-maximum distribution of AUGER which favors a predominantly mixed composition of intermediate mass nuclei, if interpreted with current hadronic interaction models tuned at the LHC  \cite{augerXmaxMeas14,augerXmaxComp14}.  First, the recently finalized Auger analysis \cite{augerXmaxComp14} finds a protonic component persisting to highest energies.  Second, a mixed composition requires a very hard injection spectrum incompatible with shock acceleration \cite{shahamPiran13,Aloisio+13} and a composition at the source which has been argued to be strange and unlikely \cite{shahamPiran13}.  Third, a  predominantly proton composition is of particular interest because both Auger and TA find evidence of correlations between UHECRs above 55 EeV and the local matter distribution \cite{TAaniso14,augerAniso14}, although anisotropy {\it per se} does not exclude mixed composition, particularly for the case of a single, nearby source \cite{Piran11,rfICRC13,kfs14}.  The final and most compelling virtue of a predominantly proton composition is that it naturally explains the shape of the spectrum from below $10^{18}$ eV to the highest energies, including the observed ``dip'' structure around $10^{18.5}$ eV and the cutoff at highest energies, without needing an ad-hoc and fine-tuned transitional component between Galactic and extragalactic cosmic rays \cite{berezGGdip05,AlBerezGaz12} with additional parameters to tune the composition and maximum energy by hand.  

The reader might also ask whether it is legitimate to set aside inferences on composition from current hadronic interaction models; the answer is ``yes''.  The nucleon-nucleon CM energy in the collision of a $10^{18}$ eV proton with the air -- 140 TeV -- is a factor 20 above current the LHC CM energy, so that the models must be extrapolated far into uncharted territory.  Furthermore, detailed comparisons by the Auger collaboration of the model predictions with a variety of observed shower properties reveals several discrepancies, including that the models underpredict the muon content of the ground shower by 30\% or more \cite{ICRC13topdown,augerHorizMuons14}, and that the model which does the best with respect to the muons at ground level (EPOS-LHC) is in the most serious contradiction with the observed depth of muon production in the atmosphere \cite{augerMPD14}.

\section*{Summary}
  

To conclude, we have shown that a scenario in which UHECRs are predominantly or purely protons can be realized, with acceleration occurring in transient AGN-like jets created in stellar tidal disruption events.  A well-studied example of such a TDE-jet, \swift\ J1644+57, displays inner and outer emission sites in which collisionless shocks satisfy the Hillas criteria. 
Thus we propose that, like in AGN models and GRB models, the  basic shock acceleration mechanism is applicable for UHECR acceleration in TDE jets.  As shown in \cite{fg09}, the conditions in such jets are such that the radiation fields within the outflow are not large enough to cool the UHECRs before they escape.  Thus both the outer and inner emission regions in TDEs may in principle be viable UHECR sources.  

We also investigated whether the total observed flux of UHECRs is compatible with the UHECR injection rate that can be expected for TDE jets; although a more thorough theoretical understanding of the UHECR acceleration mechanism is needed for a definitive conclusion, present evidence indicates the energetics are satisfactory.  Finally, we showed that the effective number of sources predicted in the protonic-UHECRs-from-TDE-jets scenario, is compatible with the even the most stringent Auger bound, i.e., the case that typical deflections are less than $3^{\circ}$.

Unlike for GRBs, the TDE-jet model for UHECR production cannot be tested directly by association of an observed transient event with a signature of UHECRs.   In the case of GRBs, prompt neutrinos are produced via photoproduction of charged pions in the source, which arrive approximately simultaneously with the gammas.  (The UHECRs themselves arrive 10's or 100's of thousands of years after the gammas or neutrinos, due to the magnetic deflections discussed previously.)  By contrast, the level of prompt neutrino production in a TDE-jet is much lower, because the radiation field in a TDE jet is less, inhibiting photopion production.   Moreover the duration of UHECR production in a TDE jet is weeks or months, so even the prompt neutrinos are broadly spread in arrival times. 

The conjecture of predominantly protonic composition, the role of transients in UHECR production, and the TDE-jet model can be tested purely observationally, as follows.  \\
\noindent $\bullet$  Whether UHECRs are protons or nuclei can in principle be determined without relying on hadronic interaction models to infer composition, by detecting or placing sufficiently strong limits on VHE photons and neutrinos produced during the propagation, as their spectra distinguish UHECR protons from nuclei.   If UHECRs are predominantly protonic, as shown above (updating earlier arguments \cite{fg09,WL09,MuraseTakami09}) their primary sources must be transients, with TDE jets the leading candidate.  
\\
\noindent $\bullet$  Whether sources are continuous or transient can be determined from the spectrum of UHECRs from a single source, because UHECRs arriving at a given epoch from a transient have similar values of rigidity $\equiv E/Z$ rather than displaying a power-law spectrum \cite{waxmanME}, c.f., Eq. (\ref{tau}).  
\\
\noindent $\bullet$  
If sources are confirmed to be transients, the presence or absence of VHE neutrinos accompanying the transient EM outburst, will distinguish between GRB and TDEs being the sources.  
\\
\noindent $\bullet$   As far as is presently known, the galaxies hosting TDEs are generally representative of all galaxies; if so, UHECR arrival directions would correlate (only) with the large scale structure, after taking into account Galactic and extragalactic magnetic deflections.  However if the rate of TDEs with jets is enhanced in active galactic nuclei, as conjectured in \cite{fg09}, an enhanced correlation of UHECRs with AGNs relative to random galaxies could potentially be seen.

\noindent{\bf Acknowledgements:}
We thank Rodolfo Barniol-Duran  for discussions and for help preparing the figure, and Kohta Murase for helpful discussions.  The research of GRF was supported in part by NSF-PHY-1212538; she thanks the Racah Institute for its hospitality during the initial stages of this work.  GRF is a member of the Pierre Auger Collaboration and acknowledges valuable interactions with Auger colleagues.  The research of TP was supported by the ERC advanced research grant ``GRBs''  the  I-CORE (grant No 1829/12) and a grant from the Israel Space Agency SELA; he thanks the 
Lagrange institute de Paris for hospitality while this research was concluded. 

\section{Appendix:  Details of \swift\ J1644+57 Analyses}

\noindent {\bf The Radio Emitting Region:}\\
Radio observations of Sw1644 began 0.9 days after the onset of the trigger \cite{Zauderer+11}  and lasted for about $ 600$ days \cite{Berger+12,Zauderer+13}. Wide frequency coverage began at 5 days.  At that time, the peak of the spectral energy distribution (SED) was at  $ \nu_p \approx 345$ GHz, with a peak flux  $F_p = 35$ mJy. The peak frequency and flux decreased  to $\sim 5$ GHz and $0.5$ mJy at 570 days.   We begin by discussing the classical equipartition method of interpreting radio observations \cite{Pacholczyk1970,ScottReadhead77,Chevalier98,kumarNarayan09}, and its relativistic generalization \cite{Barniol+13a}.  A direct application to the Sw1644 observations \cite{Zauderer+11} is likely overly naive, as we discuss subsequently.

The radio emitting region is characterized by four unknowns:  the size, $R$,   the magnetic field, $B$,  the total number of emitting electrons, $N$, and their typical Lorentz factor, $\gamma_e$. The total energy of the emitting region is the sum of the electrons' energy  $N m_e c^2 \gamma_e$ and the magnetic field energy $B^2 R^3 /6$, the baryons' energy being unimportant in the Newtonian case. 
Identifying the spectral peak as the self absorption frequency 
and using the standard expressions for the synchrotron frequency,  synchrotron  flux and  the self-absorbed flux (see e.g. \cite{Barniol+13a} for details) one can eliminate 3 of the 4 parameters and express, e.g., $B$, $N$ and $\gamma_e$ in terms of  $R$ and the observables $\nu_p$, $F_p$ and $z$. One then 
obtains  $ E = C_1 /R_{17}^6 + C_2 R_{17}^{11}$, where the first term is the electrons' energy and the second the magnetic field energy.  The constants $C_1$ and $C_2$ are given in term of the observables: 
$C_1 = 4.4 \times 10^{50} \,{\rm erg} \, ( F_{p}^4 \, d_{28}^8 \, \nu_{p,10}^{-7} \, \, (1+z)^{-11} ) $ and 
$C_2 = 2.1 \times 10^{46} \,{\rm erg} \, (  F_{p}^{-4} \, d_{28}^{-8} \, \nu_{p,10}^{10} \, (1+z)^{14}  ), $ where $d_{28}(z)$ is the   luminosity distance in units of  $10^{28}$cm and $F_p$ is the peak flux measured in mJy. 

The energy is minimized when the electrons' energy is roughly equal
to the magnetic energy, or put differently, when the system is in equipartition. 
The size of the system is strongly constrained, as the energy is a very steep function of $R$ both above and below the minimum.   
As we have three equations and four unknowns we can choose a different independent variable. For our purpose  $B R$ is most suitable and in this case one obtains $ E = \tilde C_1 /( B R)^{6/5}  + \tilde C_2 (B R)^{11/5}$.  This dependence is less steep and hence the resulting value for $BR$ is less constrained by these considerations.  If $E$ is an order of magnitude above the minimal value, $B R$ can be a factor-10 lower or a factor-3 higher than at equipartition. 

When applying the equipartition considerations to Sw1644 one has to take into account that the outflow is relativistic 
in this case. The  relativistic equipartition estimates are somewhat more complicated than the
Newtonian ones. A detailed equipartiton formalism for relativistic outflows was  developed recently by \cite{Barniol+13a}.
Like in the Newtonian case, the total energy depends very steeply on $R$, as
$ E = \hat C_1 /R^6 + \hat C_2 R^{11}$, where  $\hat C_1$ and $\hat C_2$  depend  on the observed quantities
but now also on  the outflow Lorentz factor, $\Gamma$, and on the specific geometry of the emitting region (see \cite{Barniol+13a} for details). 
Note that here the kinetic energy of the baryons within the relativistic outflow should also be included in the total energy of the system.  The bulk Lorentz factor, $\Gamma$,  can be determined using time of arrival arguments; the geometrical factors involved have to be guessed.   Given the very steep dependence of the total energy on $R$, $R$ is still  well constrained by the energy-minimization, equipartition considerations. Using these arguments \cite{Barniol+13b} find that  for the most reasonable geometry -- a narrow jet with an opening angle 0.1 -- the equipartition value of $B R$ in Sw1644 is slightly larger than $10^{17}$ Gauss cm 
(see Fig. \ref{fig:J1644_radio}).  Given 
that the equipartition estimates give a lower limit on the energy (that assumes maximal efficiency), this can be considered compatible with the Hillas condition.  Furthermore and independently, at the time of the last observations the minimal (equipartition)  energy is  $0.8 \times 10^{51}$ ergs \cite{Barniol+13b}, which is consistent with the energy required to account for the observed UHECR spectrum given the rate of TDEs, estimated above.

However the naive equipartition analysis can be doubted, since applying it leads to the conclusion that the energy of the jet increases by about a factor of 20 from the initial observations at around 5 days, to the final observations.  The apparent increase required in the jet energy appears in other approaches as well, as noticed first
by \cite{Berger+12}, who analyzed the data based on GRB afterglow modeling. This interpretation  requires an energy supply to the radio emitting region. Such an energy supply is inconsistent with the continous  decrease in the X-ray luminosity during this period \cite{Berger+12}, which supposedly reflects the activity of the inner engine and the accretion rate.  

Kumar et al.  \cite{Kumar+13} suggested that the puzzling behavior comes about because the X-ray jet passes through the radio emitting region, causing the radio emitting electrons to be continuously cooled via Inverse Compton (IC) scattering with these X-ray photons.   This efficient IC cooling decreases the observed synchrotron radio flux relative to the equipartition estimate, causing the equipartition analysis to yield a lower energy than the true energy content of the system, resulting in an erroneously-low inferred value for $B R$.  At later times  the X-ray flux diminishes, the IC cooling ceases, and the synchrotron flux increases, consistent with the late-time observations and obviating the need for an increase of the 
energy of the jet.  \cite{Kumar+13} estimate  $B R$ to be larger by a factor of a few  than the simple equipartition estimate 
\cite{Barniol+13b}. The results from their analysis are shown as the solid line in  Fig. \ref{fig:J1644_radio}; at early times $B R \approx 3 \times 10^{17}$ Gauss cm.  This is just  the value needed to accelerate $10^{20}$ eV UHECRs.

To summarize, even with naive application of equipartition, the values of the total energy and $B R$ within the radio emitting region of TDE J1644 are marginally compatible with those needed to accelerate UHECRs.  Given 
that the equipartition estimates give a lower limit on the energy (i.e., assumes maximal efficiency), and that the more physically satisfactory modeling of \cite{Kumar+13} yields a larger estimate comfortably compatible with accelerating protons to UHE, we conclude that the outer region of the Sw 1644 TDE jet likely has conditions for UHECR acceleration. 

\noindent{\bf The X-ray emitting region:} \\
The isotropic equivalent X-ray luminosity is $\approx 10^{48}$ ergs/sec.   If this were coming from the accretion disk, then by the argument of \cite{fg09} we would conclude that Eq. (\ref{lumi}) was easily satisfied within the jet.  However since in \swift\ J1644 we are viewing the jet close to its axis, there can be additional contributions to the X-ray emission, which must be modeled before $B R$ in the inner jet region can be inferred.   Unfortunately, the X-ray  observations are much less constraining on the conditions within the X-ray emitting regions than are the radio observations on the conditions in the outer region.  The X-ray spectrum is a power law;  if combined with the NIR observation it yields  a steep slope (steeper than 1/3).  The  Fermi upper limits on the GeV emission suggest a suppression of the high energy emission due to photon-photon opacity. Overall, only a single component has been observed, with no clear evidence of the peak frequency and only an upper limit on the high energy IC component. Therefore there is  significant  freedom in modeling this emission. 

Ref. \cite{Burrows+11}  models the emission (following the blazer emission model of \cite{GhiselliniTavecchio09}) as arising from a relativistic blob of plasma. They  
have put forward three models for this spectrum. 
The models differ in the dominant energy of the jet (Poynting flux or baryonic), the emission mechanims (synchrotron or external IC), the position of the emission region and the relative contributions of the disk and the jet. 
According to these models the X-rays are generated between $10^{14}$ and $10^{16}$ cm from the black hole. Estimates of $BR$ range from a few $\times 10^{15}$ for model 3 to $5 \times 10^{17}$ for model 2.

\def\apj{Astrophys.\ J.}
\def\nat{Nature}
\def\apjl{Astrophys.\ J. Lett.}
\def\apj{Astrophys.\ J.}
\def\aap{Astron.\ Astrophys.}
\def\prd{Phys. Rev. D}
\def\physrep{Phys.\ Rep.}
\def\mnras{Month. Not. RAS }
\def\araa{Annual Rev. Astron. \& Astrophys.}
\def\aapr{Astron. \& Astrophys. Rev.}
\def\aj{Astronom. J.}
\def\jcap{JCAP}

\bibliographystyle{apsrev4-1.bst}
\bibliography{UHECR_TDE,CR,flares,AGN}

\end{document}